\documentclass[12pt,preprint]{aastex}

\shorttitle{Radio pulsars as progenitors of AXPs and SGRs}
\shortauthors{Lin, \& Zhang}

\begin{document}

\title{Radio pulsars as progenitors of AXPs and SGRs: magnetic field evolution through pulsar glitches}

\author{J. R. Lin\altaffilmark{1} and S. N. Zhang\altaffilmark{1,2,3,4}}
\affil{\altaffilmark{1}Physics Department and Center for
Astrophysics, Tsinghua University, Beijing, 100084, China}
\affil{\altaffilmark{2}Physics Department, University of Alabama
in Huntsville, Huntsville, AL35899, USA}
\affil{\altaffilmark{3}Space Science Laboratory, NASA Marshall
Space Flight Center, SD50, Huntsville, AL35812, USA}
\affil{\altaffilmark{4}Institute of High Energy Physics, Chinese
Academy of Sciences, Beijing, China}

\begin{abstract}
Glitches are common phenomena in pulsars. After each glitch, there
is often a permanent increase in the pulsar's spin-down rate.
Therefore a pulsar's present spin-down rate may be much higher
than its initial value and the characteristic age of a pulsar
based on its present spin-down rate and period may be shorter than
than its true age. At the same time, the permanent increase of its
spin-down rate implies that the pulsar's surface magnetic field is
increased after each glitch. Consequently after many glitches some
radio pulsars may evolve into the magnetars, i.e., strongly
magnetized and slowly rotating neutron stars.
\end{abstract}

 \maketitle
\section{Introduction}
Pulsars are now accepted to be rapidly rotating and highly
magnetized neutron stars. The surface magnetic field of a neutron
star may be estimated from the observed period and period
derivative, i.e., $B\approx3.3\times10^{19}\sqrt{P\dot{P}}$ (G),
if we assume that the pulsar's spin-down energy is overwhelmingly
consumed by magnetic dipole radiation, the so-called
magnetic-braking model \cite{pac68}.

By the same method, we can also find a distinctive group of
pulsars with very high magnetic fields ($10^{14}$- $10^{15}$ G) in
the pulsars' P - B relation diagram. These pulsars so called
magnetars because of their extremely high magnetic fields.
Observationally these pulsars may appear as Soft Gamma-ray
Repeaters (SGRs) and Anomalous X-ray Pulsars (AXPs), in which the
steady X-ray luminosity is powered by consuming their magnetic
field decay energy \cite{dun92,pac92,kou98,thom96,hur00}. The
magnetic field decay also heats the neutron star surface that
emits thermal radiation in the X-ray band \cite{thom96}. Thus AXPs
and SGRs are a small class of pulsars with long periods (5 - 12
s), high spin-down rates and soft X-ray spectra (see Mereghetti
1999 for a comprehensive review). The AXPs and SGRs have young
spin-down ages of $10^3$ - $10^5$ years. Some of them are claimed
to be associated with some young supernova remnants (typically
$10^3$ - $10^4$ years old), showing that they may be young
pulsars; however the association of these cases are still
controversial \cite{gae01,dun02}.

In the standard model, pulsars are born with significantly
different parameters, and the magnetic fields of typical radio
pulsars remain constant or decay slowly during their lifetimes.
Thus the radio pulsars will evolve into an ``island" which gathers
most of the old pulsars. For the magnetars, very high initial
dipole fields are required to slow-down the neutron star to the
presently long periods within a relatively short time (~$10^4$
yrs) when they still have very high magnetic fields. Therefore
both the beginning and the ending of typical radio pulsars and the
magnetars are very different.

However if we simply assume that all pulsars were born with
similar initial parameters and their surface magnetic fields did
not change or decay slightly during their subsequent spin-down
lives, all observed pulsars should show an overall
anti-correlation between their period and spin-down rate. However,
in Fig.1, the pulsars (including AXPs and SGRs) with longer spin
periods tend to have higher surface magnetic field, as inferred
from $B\approx3.3\times10^{19}\sqrt{P\dot{P}}$. Therefore we are
forced to the conclusion that either all pulsars were born
significantly differently, or they were born similarly but their
surface magnetic fields have been increased during their spin-down
lives, resulting in a positive correlation between their surface
magnetic fields and spin periods.

Despite of their distinctive features, AXPs and SGRs shows many
similarities with typical radio pulsars, including the properties
of glitches. Pulsar glitches (sudden frequency jumps of a
magnitude $\frac{\Delta\Omega}{\Omega}\sim 10^{-9}$ to $10^{-6}$,
accompanied by the jumps of the spin-down rate with a of magnitude
$\frac{\Delta\dot{\Omega}}{\dot{\Omega}}\sim 10^{-3}$ to
$10^{-2}$) are a common phenomenon. A relaxation usually happens
after a glitch. However neither the period nor the spin-down rate
rate are completely recovered. For example, both the Crab pulsar
and the Vela pulsar were found to have a slow increase in their
spin-down rates and thus magnetic field increase during the last
tens of years \cite{smi99}. Observations show that the glitches
happened in the AXPs might cause the huge permanent changes to
their spin- down rates \cite{oss03,kas03}. This ``unhealed change"
in the spin-down rate might be due to the expelled magnetic field
from the core to the surface after each glitch, increasing the
surface magnetic field of a pulsar \cite{rud98}. The model of
Ruderman {\it et al.} (1998) predicts a certain relation between
the glitch rate and the spin-down age for a pulsar. Therefore with
the observed glitch parameters and the glitch rate, the long term
evolutions of pulsars caused by pulsars' glitches can be
calculated. The similarities between AXPs, SGRs and typical radio
pulsars make it reasonable to consider the possibility for some
typical radio pulsars to evolve to the magnetars, while most other
radio pulsars evolve to the ``island". In light of the discovery
of some normal radio pulsars with long periods and high magnetic
fields (comparable to the magnetars) in the Parkes multibeam
pulsar survey \cite{hob04}, it is natural to consider the
intrinsic connections between the magnetars and the radio pulsars
(especially those with high magnetic field) as an alternative to
the previously proposed model \cite{zha00}.

\section{Long term evolution of pulsars caused by glitches}

In the magnetic-braking model, assuming that the initial period of
a pulsar is much smaller than its present value, a pulsar's age
may be estimated by its characteristic spin-down age
$T_{s}=\frac{P}{(n-1)\dot{P}}$, where $n$ is the braking index and
for the dipole radiation $n=3$. However for some pulsars, their
characteristic ages are much shorter than the ages of the
associated supernova remnants. A famous example is PSR B1757-24, a
typical radio pulsar with $P=0.125$ s and $\dot{P}=1.28 \times
10^{-13}$ s s$^{-1}$, its characteristic age is 16000 yrs for the
braking index $n=3$.  However the associated supernova remnant SNR
G5.4-1.2 is believed to be produced between 39000 to 170000 yrs
ago \cite{fra91,gae00,man02}. This pulsar was reported to have a
very small proper motion in this sky \cite{tho02}, indicating that
either this pulsar should be very old and its magnetic field
increased in its history \cite{tho02}, or this association is
wrong. Recently there are some other models supporting the PSR
B1757-24 and G5.4-1.2 association \cite{gva04}.

Therefore discrepancy between the spin-down age and associated
supernova age for PSR B1757-24 appears to be real. This large
discrepancy can not be explained by simply involving a smaller
braking index, which in this case would require n$<$1.2, in
contrast to the smallest braking index of n=1.4 (for the Vela
pulsar) ever known for all pulsars. However, implied from the
measured $\Omega$, $\dot{\Omega}$ and $\ddot{\Omega}$ of this
pulsar, the braking index for PSR B1757-24 is $3 - 30$
\cite{lyn96}. A fall-back disk model was proposed to explain the
age discrepancy \cite{mar01,shi03}. However a pulsar in a
propeller phase should produce a dim thermal x-ray emission,
contrary to the observed bright non-thermal emission which is
consistent with the standard magnetospheric emission model
\cite{kas01}. Alternatively, the pulsar glitches may be the source
of the above age discrepancy. Usually after a glitch, both the
period and the spin-down rate of the pulsar are changed, though
the period change is usually negligible. However the accumulated
increase in the spin-down rate after many glitches will cause a
underestimation to the pulsar's age.

The possibility for the magnetic growth \cite{bla83} and to
reconcile the age discrepancy of some pulsars by the magnetic
field growth \cite{cha95} or pulsars' glitches \cite{mar04} have
been discussed previously. With the data of the observed glitch
parameters of some pulsars, we can do quantitatively more detailed
calculations. In this work we investigate the roles of pulsar's
glitches in the long term evolution of pulsars, in order to
explain the observed positive correlation between pulsars' surface
magnetic fields and their periods, and the large discrepancy
between pulsars' characteristic ages and their associated
supernova remnants.

We take PSR B1757-24 as an example to illustrate the long term
evolutionary effects caused by pulsars' glitches. A giant glitch
in PSR B1757-24 was reported \cite{lyn96}. The long-term
post-glitch relaxation fit shows that
$\frac{\Delta\dot{P}_p}{\dot{P}}\approx 0.0037$, where
$\Delta\dot{P}_p$ is the permanent change to $\dot{P}$ after the
glitch. Based on the observational fact that different pulsars
have on the average distinctive $\frac{\Delta\dot{P}_p}{\dot{P}}$,
it is not unreasonable to assume that similar giant glitches have
happened repeatedly in the history of PSR B1757-24, therefore we
can describe its spin-down history by the following three
equations:
\begin{equation}
P-P_0=\int^{\tau}_{0}\dot{P}(t)\mathbf{d}t=\sum\int^{
\tau_{n}}_{0}\dot{P}_{n}(t)\mathbf{d}t.
\end{equation}
where $\tau_{n}$ is the interval between two adjacent glitches.
The previous observations indicate that the glitch activity may be
negatively related to the pulsar's spin-down age except for the
youngest pulsars such as the Crab pulsar \cite{lyn95}. Both this
relation and the exception can be well explained by a theoretical
work in which $\tau_{n}\propto\frac{P}{\dot{P}}\propto T_{s}$ for
a pulsar \cite{rud98}. We adopt this relation in our calculations.
Assuming that between two adjacent glitches, the surface magnetic
field of the pulsar remains constant, we have
\begin{equation}\dot{P}_n(t) P_{n}(t)=\textbf{Const}_{n}\end{equation}
For PSR B1757-24 we assume the following relationship is true for
every glitch:
\begin{equation}\dot{P}_{n+1}(\tau_{n})=\alpha\dot{P}_{n}(0),\end{equation}
where $\alpha=1.0037$ for PSR B1757-24.

%\clearpage

%%%%%%%%%%%%%%%%%%% FIGURE 1%%%%%%%%%%%%%%%%%%%%%%%%%%%%
\begin{figure*}[htb]
\centerline{\includegraphics[scale=0.8]{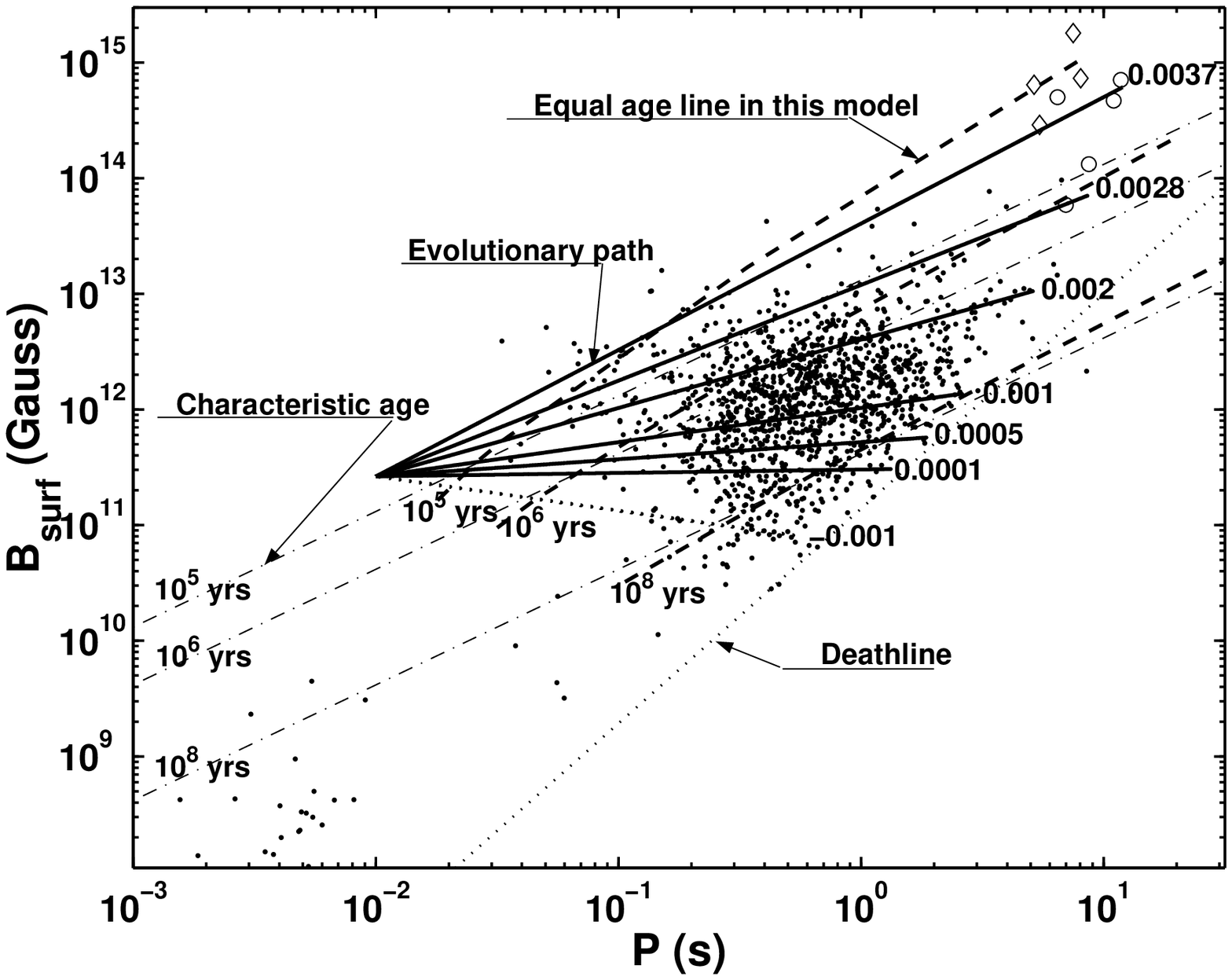}}\label{fig:fig1}
\caption {All known isolated pulsars are shown in this figure.
Filled circles are millisecond and ``regular" radio pulsars, open
circles are AXPs, and diamonds are SGRs. The pulsars with longer
spin periods tend to have higher surface magnetic field, as
inferred from $B\approx3.3\times10^{19}\sqrt{P\dot{P}}$. The
dotted-line is the ``death-line" for radio pulsars \cite{che93}.
The magnetic field evolution of pulsars are caused by the
permanent changes to the spin-down rates after glitches. Different
solid lines denote different values of $\frac{\Delta\dot
P_{p}}{\dot P}$. The line for $\frac{\Delta\dot P_p}{\dot
P}=0.0037$ is the evolutionary path of PSR B1757-24, calculated
from equations (1), (2) and (3). Assuming that all pulsars were
born with the same initial surface magnetic field and spin period,
but different glitch properties, their different evolutionary
paths are also shown for different values of $\frac{\Delta\dot
P_p}{\dot P}$. Pulsars on the same dashed-lines have the same age
as calculated in our model, in contrast to the characteristic ages
(dashed-dotted lines) of pulsars based simply on their present day
period and spin-down rate without taking into account of pulsar
glitches.}
\end{figure*}

%\clearpage

%%%%%%%%%%%%%%%%%%% FIGURE 1%%%%%%%%%%%%%%%%%%%%%%%%%%%%
Taking its present values of $P=0.125$ s and $\dot{P_n}=1.28
\times 10^{-13}$ s s$^{-1}$ and assuming that its true age is
$\tau=10^5$ yrs and its initial period is $P_0=10$ ms, we get
$n=1495$ from the above equations and its initial surface magnetic
field is about $2.6\times 10^{11}$ Gauss. We can also estimate
that its glitch rate would be about once per 3.4 years when it was
1000 years old, similar with the observed glitch rate for the Crab
pulsar. If we assume that future glitch rate for PSR B1757-24 will
continue to follow this pattern, then after $2 \times 10^5$ years,
its characteristics will be similar to AXPs as shown in Fig.1.
Assuming that all pulsars were born with the same surface magnetic
field and spin period, but different glitch properties, their
different evolutionary paths are also shown in Fig.1 for different
values of $\frac{\Delta\dot P_p}{\dot P}$. The equal-age lines are
also shown in the figure, in contrast to the characteristic ages
of pulsars based simply on their present day period and spin-down
rate without taking into account of pulsar glitches. Under the
same assumption, in Fig.2 we show the estimated ages for the
pulsars with given $P$ and $\dot{P}$.

%\clearpage

%%%%%%%%%%%%%%%%%%% FIGURE 1%%%%%%%%%%%%%%%%%%%%%%%%%%%%
\begin{figure}[h]
\label{fig:fig2} \centerline{\plotone{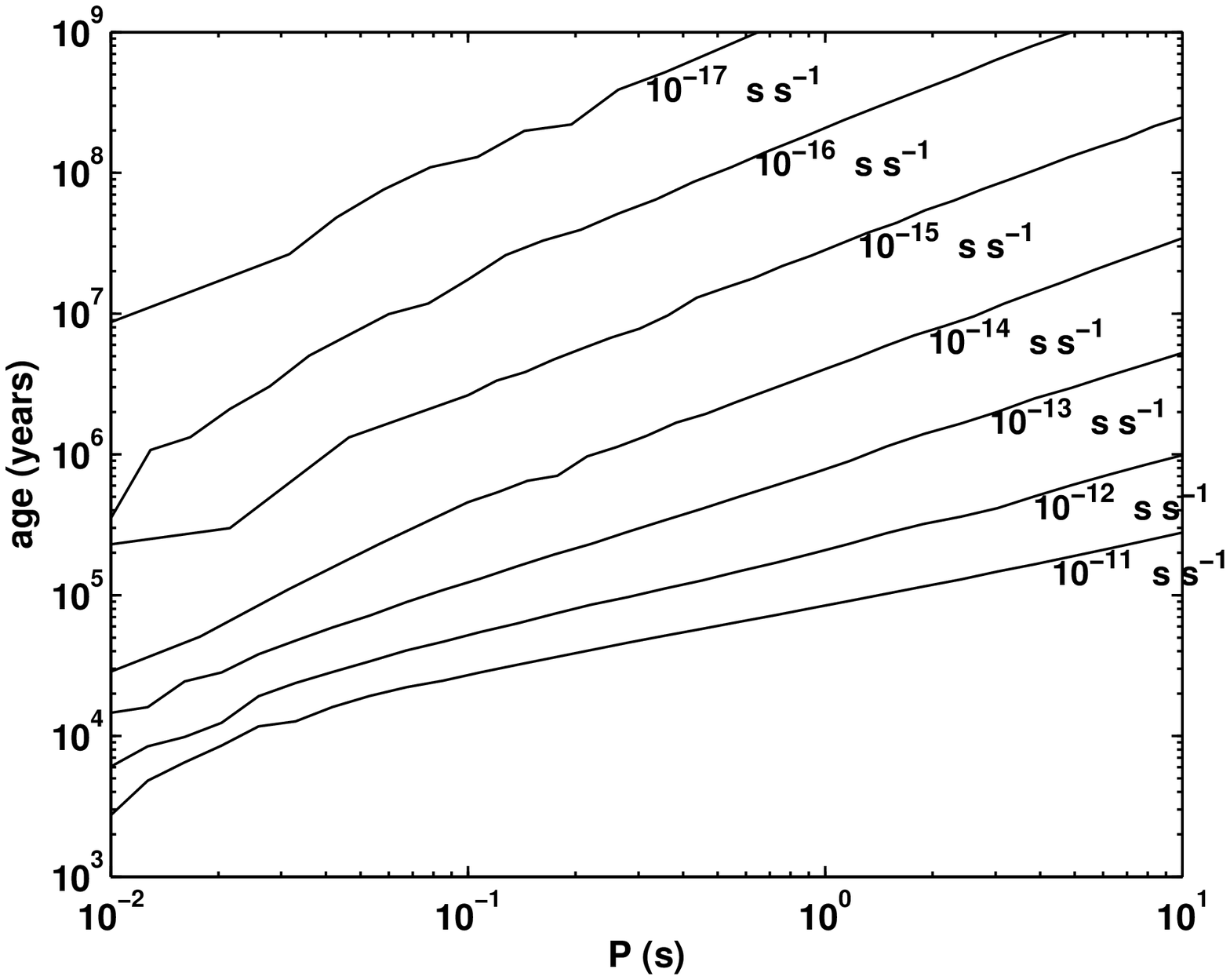}}\caption {Assuming
the same initial condition, for given $P$ and $\dot{P}$ we can
calculate the pulsar's age from equations (1), (2) and (3). Since
the effect caused by the glitches is a accumulated process, our
model for pulsar's age estimate is reliable only for pulsars older
than $10^5$ years, but  has considerable uncertainties for younger
pulsars. The solid lines are for the different values of
$\dot{P}$.}
\end{figure}
%%%%%%%%%%%%%%%%%%% FIGURE 1%%%%%%%%%%%%%%%%%%%%%%%%%%%%

%\clearpage

\section{Conclusions and discussion}

Based on our results presented above, we have the following
conclusions and remarks:

(1) Pulsar glitches, especially the permanent changes to their
spin-down rates, are important for the long term evolutions of
pulsars. The previous evolutionary history and future ``fate" of a
pulsar may be calculated within the frame work of the
magnetic-braking model, if its glitch properties are known.

(2) If we assume all pulsars were born similarly, then the
positive $P - B$ correlation may be explained naturally. However
any slight differences in the initial conditions for different
pulsars may cause a large uncertainties for their age estimates
when pulsars are younger than 10$^5$ years , as seen in Fig.1 and
Fig.2; the age estimated from our model is reliable only for those
pulsars older than 10$^5$ years, such as PSR B1757-24.

(3) Our model suggests that some radio pulsars, such as PSR
B1757-24 and other pulsars which exhibit large values of
$\frac{\Delta\dot{P}_p}{\dot{P}}$ and are thus along similar
evolutionary paths shown in Fig. 1, may eventually evolve into
AXPs and SGRs within 10$^5$ to 10$^6$ years after their birth,
contrary to the classical ``magnetar" model in which they are very
young neutron stars born with very high magnetic fields
\cite{dun92}. Our model requires that AXPs and SGRs have glitches
with large values of $\frac{\Delta\dot{P}_p}{\dot{P}}$ and a
glitch rate of once per several years; this is consistent with the
observed glitch properties of AXPs \cite{oss03,kas03}. However, if
the associations of young SNRs with some magnetars are true
(however see \cite{gae01} for arguments against the associations),
we can not rule out the possibility that some magnetars might be
born with ultra-high magnetic field, or their glitch histories are
significantly different from known radio pulsars.

(4) In our model, the small number of known magnetars compared to
``regular" radio pulsars requires their progenitors should also be
rare. This is roughly consistent with the small number of pulsars
along the evolutionary paths leading to the magnetars, as can be
seen in Fig.1. Since the values of
$\frac{\Delta\dot{P}_p}{\dot{P}}$ decide the pulsars' fates, we
can estimate the expected percentage of magnetars with Fig.1. In
Fig.1, only the pulsars with $\frac{\Delta\dot{P}_p}{\dot{P}}>
0.0028$ would evolve to the magnetars. So among all these isolated
pulsars in this diagram, $4.6\%$ of them will evolve into
magnetars. In the P-B diagram, there is an observational region
for the AXPs and the SGRs. Estimation based on our model shows
that the time scale for the pulsars to go through this region is
$1.5\times 10^4$ yrs and the total lifetime (from the initial
point we set in our model to the end of the magnetar phase) for
PSR B1757-24 is $2\times 10^5$ yrs. Therefore, under the
assumption of a uniform pulsar birth rate over time, for the
pulsars that will evolve into magnetars, around $7.5\%$ of their
lives will be in the magnetar phase. Currently it is difficult to
compare accurately the expected number of magnetars with known
magnetars, because the samples for both radio pulsars along the
evolutionary path and magnetars may be quite incomplete.

(5) Our model does not include the long term magnetic field decay
(MFD) of pulsars \cite{gun70}. However for pulsars with active
glitches, the magnetic field increase by glitches overwhelms the
slow magnetic field decay. In the case that the significant
magnetic field decay is inevitable such as the magnetars whose
X-ray emission is believed to be powered by the magnetic field
decay energy, our model infers a time scale of only $3 \times
10^4$ years for the magnetic field to increase from 10$^{14}$
Gauss to 10$^{15}$ Gauss. In contrast, the estimated time scale of
MFD (induced by the Hall cascade) from 10$^{14}$ Gauss to
10$^{13}$ Gauss is 10$^5$ years, and it takes more than 10$^7$
years for the same amount of decay driven by ambipolar diffusion
\cite{col00}. A more realistic model for pulsars with ``weak"
glitch properties should also include the long term magnetic field
decay process. We will investigate this in the future.

(6) In Fig.2 the ages are estimated for most pulsars according to
their period and period derivative by equations (1), (2) and (3).
These predictions may be tested with future pulsar and SNR
observations.

(7) Finally we should mention that since our model does not assume
different radiation mechanisms for all pulsars, the birth and
death lines for pulsars remain unchanged.

\noindent {\bf Acknowledgement: } We thank Drs. Zigao Dai, Yang
Chen, Renxin Xu and Guojun Qiao for valuable comments to the
manuscript. We also thank the anonymous referee whose comments
allowed us to clarify several points and improve the readability
of the manuscript substantially. This study is supported in part
by the Special Funds for Major State Basic Research Projects
(10233010) and by the National Natural Science Foundation of
China. SNZ also acknowledges supports by NASA's Marshall Space
Flight Center and through NASA's Long Term Space Astrophysics
Program.

\end{document}